\documentclass[a4paper]{article}

\usepackage{INTERSPEECH2021}
\usepackage{multirow}
\usepackage{paralist}

\title{CCATMos: Convolutional Context-aware Transformer Network for Non-intrusive Speech Quality Assessment}
\name{Yuchen Liu$^{1,2}$, Li-Chia Yang$^1$, Alex Pawlicki$^1$, Marko Stamenovic$^1$}
\address{
  $^1$Bose Corporation, USA \\
  $^2$Department of Computer Science, Indiana University, USA
  }
\email{liu477@iu.edu, richard\_yang@bose.com }

\begin{document}

\maketitle
\begin{abstract}
Speech quality assessment has been a critical component in many voice communication related applications such as telephony and online conferencing. Traditional intrusive speech quality assessment requires the clean reference of the degraded utterance to provide an accurate quality measurement. This requirement limits the usability of these methods in real-world scenarios. On the other hand, non-intrusive subjective measurement is the ``golden standard" in evaluating speech quality as human listeners can intrinsically evaluate the quality of any degraded speech with ease. In this paper, we propose a novel end-to-end model structure called Convolutional Context-Aware Transformer (CCAT) network to predict the mean opinion score (MOS) of human raters. We evaluate our model on three MOS-annotated datasets spanning multiple languages and distortion types and submit our results to the ConferencingSpeech 2022 Challenge. Our experiments show that CCAT provides promising MOS predictions compared to current state-of-art non-intrusive speech assessment models with average Pearson correlation coefficient
(PCC) increasing from 0.530 to 0.697 and average RMSE decreasing from 0.768 to 0.570 compared to the baseline model on the challenge evaluation test set.

\end{abstract}

\section{Introduction}
Evaluation of speech quality is a key component of any audio application involving voice communications including telecommunications, consumer electronics and hearing assistance. Objective intrusive speech quality assessment is widely used to evaluate speech quality from various speech processing tasks such as speech enhancement \cite{williamson2015complex, weightblockunit}, dereverbration \cite{li2021loss} and speech synthesis \cite{wang2017tacotron}. Perceptual Evaluation of Speech Quality (PESQ) \cite{recommendation2001perceptual} is designed to test the speech and voice quality of speech as perceived by human beings. Short-Time Objective Intelligibility (STOI) \cite{taal2011algorithm} focuses more on intelligibility measurement. Hearing-Aid Speech Quality Index (HASQI) \cite{kates2010hearing} is a method created specifically to evaluate speech quality for hearing aid users.

All of these objective intrusive methods require a clean reference speech signal for evaluating the quality of the given utterance. However, a clean reference is often unavailable in real-world situations. Because of this limitation, intrusive speech assessment methods are unable to be used in most real-world voice communications scenarios, such as real-time phone call quality, on-device acoustic noise suppression  and  virtual meeting speech quality assessments. 

On the other hand, objective non-intrusive speech assessment methods do not require a clean reference speech signal for the evaluation process, which increases their usability. The current objective non-intrusive specification provided by ITU-T is P.563 \cite{falk2004single}. P.563 is the first non-intrusive speech quality assessment model designed for telephony applications. It assesses the quality of speech transmitted over a narrow-band telephone channel. However, P.563 can only applied on narrow-band signals with certain length, which limits its applicability. Other attempts at object non-intrusive speech assessment methods such as ANIQUE \cite{kim2005anique}, and the speech-to-reverberation modulation energy ratio (SRMR) \cite{falk2010non} do not show enough correlation to subjective human assessment.

Recently, there has been  flurry of work aimed at making objective evaluation methods non-intrusive by applying data-driven techniques. The approach of these data-driven methods is often to pose intrusive speech quality assessment as a regression problem where the goal is to map directly from the audio to the evaluation score, eliminating the requirement of a reference signal. Quality-Net \cite{fu2018quality} uses a bidirectional long short-term memory (BLSTM) based network to predict both frame-level and utterance-level PESQ scores. Similarly, STOI-Net \cite{zezario2020stoi} predicts the STOI by using a CNN-BLSTM network. Zhang \textit{et al.} \cite{zhang2021end} implement a multi-task framework and predict mean opinion score (MOS), PESQ, the extended short-time objective intelligibility (eSTOI) and signal to distortion ratio (SDR) simultaneously with a pyramid BLSTM network.

While objective metrics are valuable, subjective speech assessment by human raters is still considered the ``golden standard" of speech quality assessment \cite{reddy2021dnsmos} since human perception of audio quality is highly subjective \cite{fedorov2020tinylstms}. Subjective assessments are often conducted by aggregating perceptual ratings from multiple raters ranging from 1 (worst) to 5 (best) into a single MOS score per utterance. In contrast to objective metrics, raters can recognize and evaluate degraded utterances easily with minimal reference. 
ITU-T P.808 \cite{rec2018p} defines a framework to  reproducibly collect MOS-annotated data by providing tools for ensuring inter-rater reliability and spam filtering. Unfortunately, ratings can be noisy due to inherent differences in human perception even when adhering to current best practices \cite{itut1401}. Moreover, subjective speech assessment is costly, complex and time consuming, severely limiting its scalability.

Due to the noisy and nonlinear nature of perceptual speech quality, deep learning-based data-driven approaches to speech quality assessment have moved to the forefront of non-intrusive speech quality research. Recent work has tackled the problem with several neural network designs. For instance, MOSNet \cite{mosnet} uses a convolutional neural network (CNN) followed by a BLSTM to predict both frame level and utterance level scores. DNSMOS \cite{reddy2021dnsmos} uses a self teaching \cite{kumar2020sequential} approach to train a CNN-based model. AutoMOS \cite{patton2016automos} combines a classification loss with a regression loss to predict more accurate MOS scores. NISQA
(v2.0) \cite{NISQA} uses a CNN to extract frame-wise features and attention model to estimate the MOS score.

In this paper, we propose a novel end-to-end non-intrusive speech assessment method which directly estimates MOS scores for varied real-world speech. The proposed model takes a raw speech signal as input and generates context-aware embeddings to capture abundant contextual information of the given speech at multiple timescales. Several convolutional layers are then used to exploit local features, followed by several transformer encoder blocks to aggregate long-range utterance level features. A MOS score is predicted for each frame and the final utterance MOS score is calculated by averaging all the predicted frame-level predictions.

The details of our proposed model are provided in section \ref{Method} and the dataset details are shown in section \ref{Datasets}. Experimental results and conclusions can be found in section \ref{results} and \ref{conclusion}, respectively.

\begin{figure*}[t]
  \centering
  \includegraphics[scale=0.38]{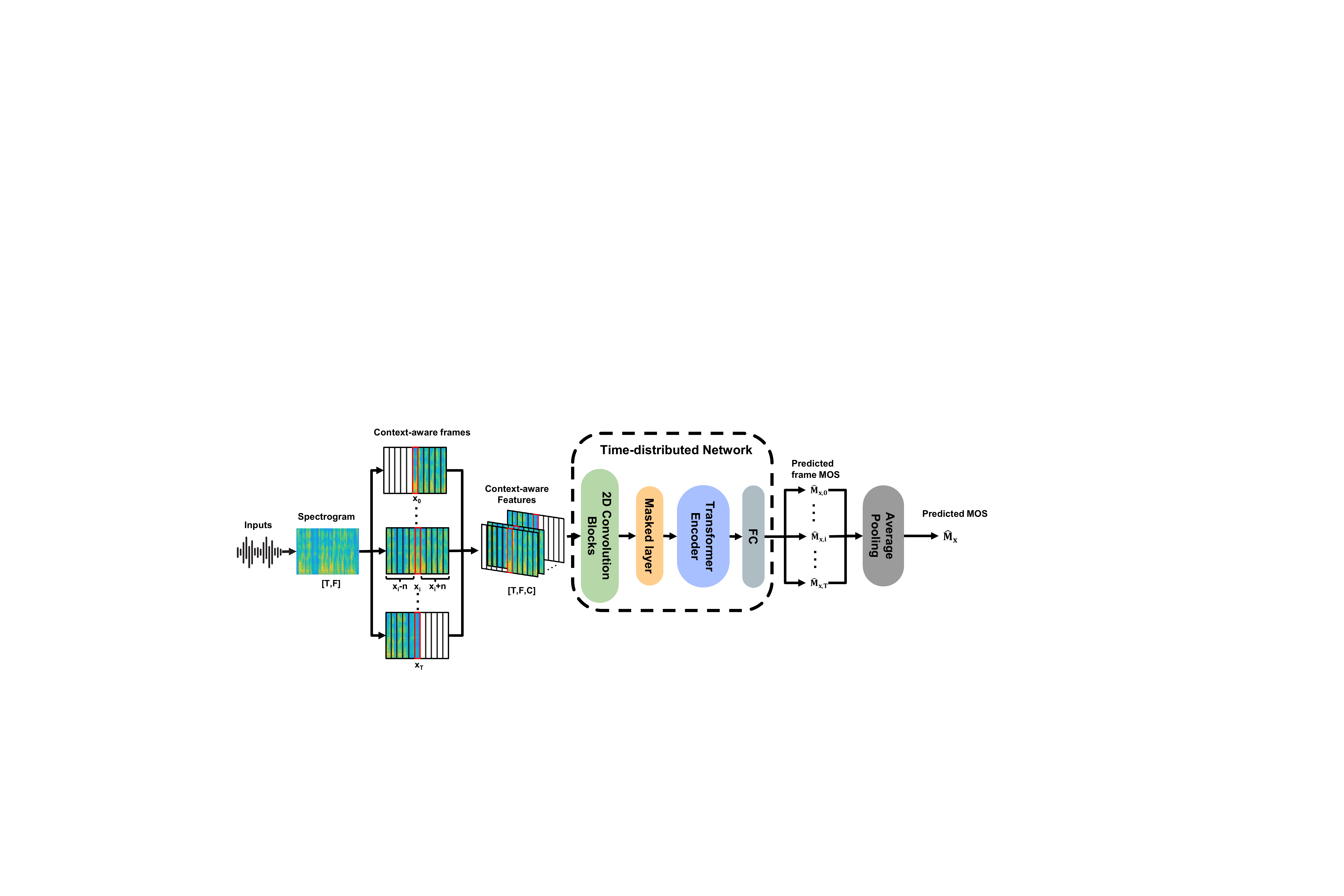}
  \caption{Overall Model Structure, the components of transformer encoder are illustrated in Figure\ref{fig:encoder}.}
  \label{fig:model}
\end{figure*}

\section{Method} \label{Method}
In this section, we describe the detailed model structure and experimental method for our MOS estimator. An illustration of the model architecture can be found in Fig. \ref{fig:model}. First, the input signal is converted into a time-frequency (T-F) domain log-scaled magnitude spectrogram. The spectrogram is then aggregated to generate context-aware features. Next, the context-aware features are fed into a time-distributed neural network with three 2D convolutional blocks, a masking layer, a transformer encoder block and several fully connected (FC) layers. A MOS score is predicted for each context-frame as well as for the overall utterance. The final utterance-level predicted MOS score is calculated by average-pooling the predicted frame-level MOS. 

\subsection{Feature Extraction}
We use the short-time Fourier transform (STFT) \cite{bracewell1986fourier}  to extract the T-F domain spectrogram for each time domain input signal. The  log-scaled magnitude spectrogram is calculated next, resulting in a signal $x \in \mathbb{R}^{T \times F}$. In order to capture longer-term contextual information in the time-distributed network, we use context-aware frames to maintain the temporal information during the training process. For each frame in the log-scaled magnitude spectrogram $x_{i}$, $n$ past frames and $n$ future frames are concatenated with the current frame $x_{i}$ to generate a context-aware frame.  Frames at the beginning and end of the signal are zero-padded where necessary. The context size $C$ can be calculated as $2n+1$. The resultant context-aware features can be expressed as a three-dimensional tensor $c \in \mathbb{R}^{T \times F  \times C}$.

\subsection{Model Structure}
The overall regression network mainly contains two components, a time-distributed network used to provide the frame-level estimated MOS score and an average pooling layer used to average the frame-level predicted MOS to the final utterance-level predicted MOS score.

The time-distributed network contains three parts: 2D convolutional blocks, transformer encoder blocks \cite{vaswani2017attention} and several fully connected (FC) layers. The 2D convolutional blocks are used to extract local contextual features. They contain three 2D convolutional layers with kernel size $[5,5]$ and stride 1. Each convolutional layer is followed by a ReLU activation function and an average pooling layer with the pool size $[2,2]$. The downsampling rate for each convolutional block is $[8,8]$, which results in an output embedding dimension of $[T,\frac{F}{8},\frac{C}{8}]$. We choose not to use a bias vector in order to allow the subsequent masking layer to stabilize training.

\begin{figure}[t]
  \centering
  \includegraphics[trim={0 1.5cm 0 0},clip,scale=0.24]{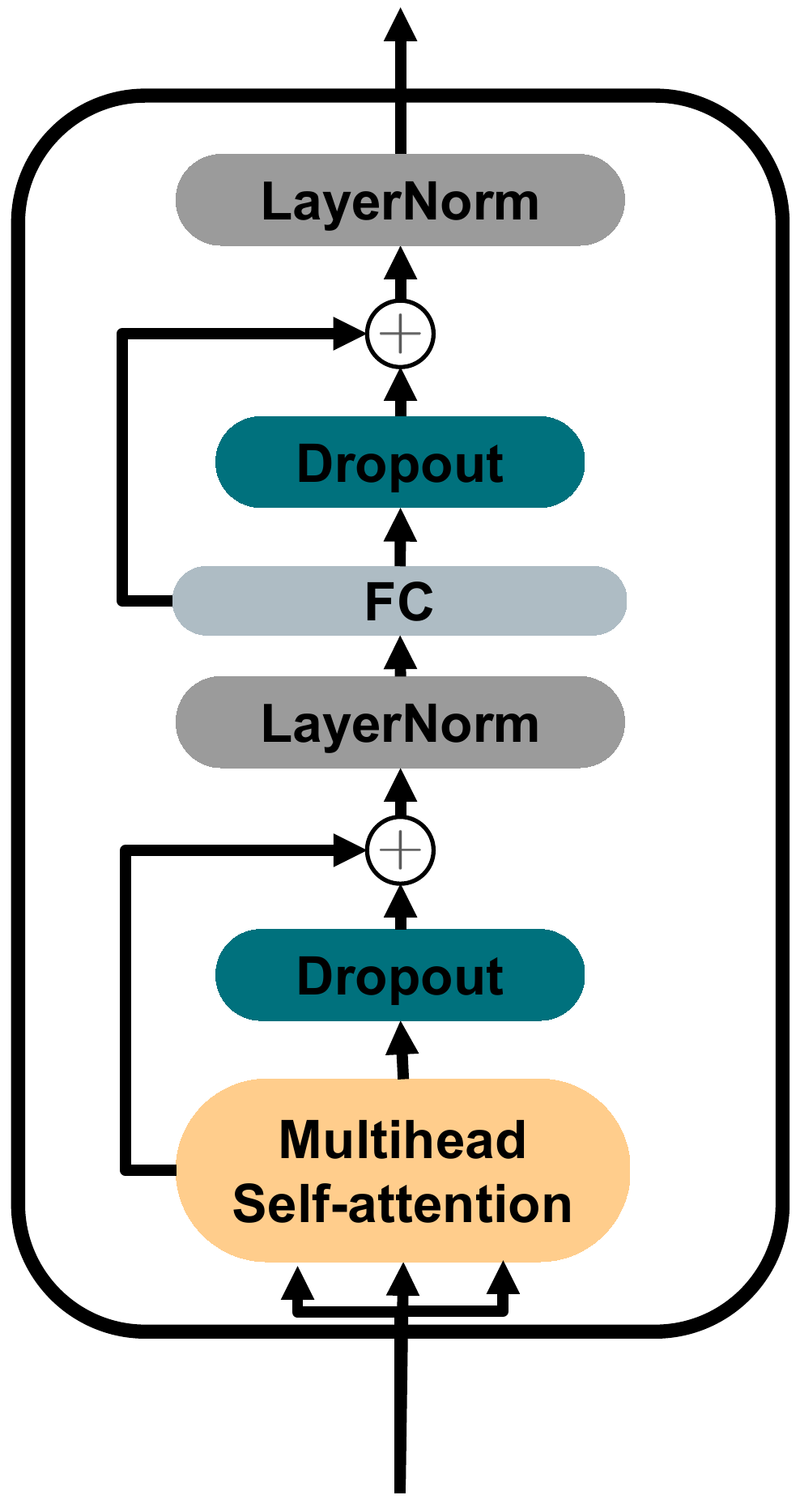}
  \caption{Transformer Encoder.}
  \label{fig:encoder}
\end{figure}

Next, the masked context-aware convolutional results are fed into several transformer encoder blocks. The transformer encoders are used to capture global contextual interactions at the utterance level. Each block consists of a multi-head self-attention layer with skip connection and a fully connected layer (illustrated in Fig.\ref{fig:encoder}). This is followed by two FC layers with ReLU activation and with a maximum value of 5 in order to predict the frame-level MOS score $\hat{M}_{x,1}...\hat{M}_{x,T}$. Finally, an average pooling layer is used to combine the frame-level MOS predictions into an utterance-level MOS. We apply dropout and L2 regularization throughout the network to prevent overfitting.

\begin{table}[]
\centering
\scriptsize
\begin{tabular}{lccc}
\hline
                 & Model 1 & Model 2 & Model 3 \\ \hline
feature          & STFT    & Mel     & Mel     \\
context size     & 11      & 11      & 11      \\
Conv filters     & 16      & 8       & 16      \\
Conv filter size & 5       & 5       & 5       \\
Transformer encoders       & 4       & 2       & 4\\
Transformer FC units     & 256     & 256     & 256     \\
Transformer Att heads        & 4       & 2       & 4       \\
FC units         & 512     & 256     & 512     \\
FC layers        & 2       & 2       & 2      \\ 
batch size        & 4       & 16       & 4      \\
dropout        & 0.15       & 0.19       & 0.15        \\ 
learning rate        & 1e-5       & 4.2e-4        & 1e-5  \\
l2 regularizer        & 6e-3       & 1e-3       & 6e-3   \\
 \hline
\end{tabular}
\caption{Network configuration for the final models. Note that Model 3 has an identical configuration with Model 1 except the input feature type, this is due to the experiment on feature variants is conducted after the hyper-parameter search.}
\label{table:params}
\end{table}

\subsection{Loss Function}
Inspired by QualityNet \cite{fu2018quality}, we use a MOS-weighted mean squared error (MSE) loss for frame-level and a regular MSE loss for utterance-level MOS prediction.

The utterance-level MOS ground truth $M_{x} $is also used as the frame-level MOS ground truth. However, the gap between utterance-level MOS score and the frame-level MOS score increases as the quality of the utterance decreases. Therefore, a conditional weight is added to the frame-level MSE to de-emphasize this bias. The conditional weights are calculated as:

\begin{equation}
\alpha\left(M_{x}\right)=10^{\left(M_{x}-5\right)}
\end{equation}

where 5 is upper-bound of the MOS score. With $X$ denoting the total number of training utterances and $T$ as the number of frames in each utterance, $Ux$, the final loss can be expressed as:
\begin{equation}\resizebox{.85\hsize}{!}{$
L=\frac{1}{X} \sum_{x=1}^{X}\left[\left(M_{x}-\hat{M}_{x}\right)^{2}+\frac{\alpha\left(M_{x}\right)}{T} \sum_{t=1}^{T(U_x)}\left(M_{x}-\hat{M}_{x, t}\right)^{2}\right]$}
\end{equation}

\subsection{Hyper-parameter Tuning and Ensemble Modeling}

We perform hyper-parameter tuning using a Bayesian optimization strategy \cite{bergstra2015hyperopt} to search for the best combination of network parameters. The parameter search space includes the following: context size $C$ for the input features; filter size and number of filters for convolutional layers; number of transformer encoders, attention heads, and feed-forward units in the transformer encoder; hidden units and number of layers for FC layers; and finally, batch size, learning rate, dropout rate, lambda for L2 regularizer for training configurations.

We implement the hyper-parameter search for 24 runs and select the top two parameter sets as our candidate networks. In addition, we train two feature variants with log-magnitude STFT spectrograms and log-magnitude mel spectrograms on each candidate network, resulting in four candidate models. Finally, we perform model ensembling by selecting three best performing models based on the evaluation results and average their output as the final 
MOS prediction. The model parameters of the three selected models are shown in Table \ref{table:params}.

\section{Datasets} \label{Datasets}
We use the dataset provided by the ConferencingSpeech 2022 challenge\footnote{https://github.com/ConferencingSpeech/ConferencingSpeech2022}. Details about the dataset can be found in \cite{yiconferencingspeech}. The full training dataset is a composite of three separate corpora. The first is the Tencent corpus, which contains 8,366 utterances from reverberant environments and 3,197 utterances from non-reverberant ones. All speech clips are in Chinese. In addition to reverberation, all of the utterances contain distortions which would regularly occur in online conferencing such as background noise, packet loss and network jitter. The MOS label is generated using a crowdsourcing method based on ITU-T P.808 which provide a MOS score from 1 to 5. The second dataset is the NISQA corpus. The NISQA corpus is an open source dataset containing 14,432 English utterances. The dataset is pre-partitioned into training, validation and test sets and each set contains utterances with both real and simulated network degradations, similar to Tencent. MOS labels are also crowdsourced in a similar way to the Tencent dataset. The last dataset is the PSTN corpus which consists of English Librivox utterances with additive noise pushed through a PSTN send to a VoIP receive endpoint. The dataset contains 58,709 utterances each with a fixed length of 10 seconds. 

Since the Tencent and PSTN corpora are released without training and development partitions, we use the following procedure to define our own: we first extract audio embeddings for each recording using OpenL3 \cite{cramer2019look}, and leverage the embedding as the audio feature for the Kennard-Stone algorithm \cite{kennard1969computer} to split 90\% of each corpus into our training set and the remaining 10\% as our development set. As for the NISQA corpus, we include the LIVETALK and FOR partition of the NISQA TEST corpus as our development set, and place the remaining NISQA data into the training set. Note that we include the P.501 portion of the NISQA TEST data in our training set to enhance the domain diversity of the training data.

The final evaluation test set provided by the challenge contains three sub-partitions, Tencent, PSTN and TUB. The Tencent test set include 2,898 Chinese speech clips with 799 clips from reverberant environments and 2099 clips from environments without reverbration. The TUB eval dataset contains 434 German and English utterances and the PSTN eval dataset contains 1,454 ten second speech clips.

\section{Experimental Results} \label{results}

In this section, we first present two sets of ablation studies to  
\begin{inparaenum}[(i)]
    \item verify the benefit of context-aware features and
    \item highlight a cross-dataset generalization error issue that we observed.
\end{inparaenum} 
We present the individual results of our final models along with the benchmarks against two recent works on our development set, and attach the ConferencingSpeech 2022 challenge results.

The challenge provides two baseline models. The models are trained on random splits of the challenge datasets where the specification of train and development split of the datasets are not available. As a result, we are unable to compare our models directly with the baselines with the available data. Hence, we benchmark the latest NISQA \cite{NISQA} model (NISQA v2.0) and the DNSMOS \cite{reddy2021dnsmos} model on our development datasets to provide a fair comparison with our proposed methods. We evaluate our results using average Pearson’s correlation coefficient (PCC) and root mean squared error (RMSE).

\begin{table*}[t]
\centering
\resizebox{\linewidth}{!}{%
\begin{tabular}{lllllcccccccc}
\hline
\multirow{2}{*}{Dataset} & \multicolumn{2}{c}{NISQA\cite{NISQA}}                          & \multicolumn{2}{c}{DNSMOS\cite{reddy2021dnsmos}  }                       & \multicolumn{2}{c}{Model 1} & \multicolumn{2}{c}{Model 2} & \multicolumn{2}{c}{Model 3} & \multicolumn{2}{c}{CCATMOS (Ensemble)} \\ \cmidrule(r){2-3}  \cmidrule(r){4-5}  \cmidrule(r){6-7}  \cmidrule(r){8-9}  \cmidrule(r){10-11}  \cmidrule(r){12-13}    
                         & \multicolumn{1}{c}{PCC} & \multicolumn{1}{c}{RMSE} & \multicolumn{1}{c}{PCC} & \multicolumn{1}{c}{RMSE} & PCC          & RMSE         & PCC          & RMSE         & PCC          & RMSE         & PCC           & RMSE         \\ \hline
Tencent Dev              & 0.719                  & 0.953                   & 0.737                   & 0.888                   & 0.872       & 0.526       & 0.874       & 0.516       & 0.882       & 0.502       & \textbf{0.901}        & \textbf{0.452}       \\
PSTN Dev                & 0.683                  & 1.071                   & 0.639                  & 0.754                   & 0.754       & 0.605       & 0.750       & 0.591       & 0.766       & 0.573       & \textbf{0.779}        & \textbf{0.559}       \\
NISQA \_TEST\_LIVETALK   &   0.778                      &       \textbf{0.617}                   &       0.580                  &          0.943                & 0.708       & 0.737       & 0.747       & 0.720       & 0.731       & 0.718        & \textbf{0.779}        & 0.646       \\
NISQA \_TEST\_FOR        &       \textbf{0.891}                  &        \textbf{0.418}                  &        0.586                 &         1.327                & 0.763       & 0.619       & 0.732        & 0.681       & 0.804       & 0.634       & 0.827        & 0.541       \\ \hline
\end{tabular}}
\caption{Results on our own division of development sets}
\label{table:result}
\end{table*}

\subsection{Ablation Study}
\begin{table}[]
\resizebox{\columnwidth}{!}{%
\begin{tabular}{lllcc}
\hline
\multirow{2}{*}{Dataset} & \multicolumn{2}{c}{w/o context}                    & \multicolumn{2}{c}{w/o PSTN Train} \\ \cmidrule(r){2-3}  \cmidrule(r){4-5} 
                         & \multicolumn{1}{c}{PCC} & \multicolumn{1}{c}{RMSE} & PCC              & RMSE            \\ \hline
Tecent Dev              & 0.858                  & 0.564                   & 0.863           & 0.544          \\
PSTN Dev                & 0.725                  & 0.650                   & 0.654           & 0.680          \\
NISQA\_TEST\_LIVETALK   & 0.698                  & 0.810                   & 0.699           & 0.767          \\
NISQA\_TEST\_FOR        & 0.756                  & 0.642                   & 0.761           & 0.588          \\ \hline
\end{tabular}}
\caption{Ablation studies}
\label{table:abtest}
\end{table}

\subsubsection{Context-aware features}

In Table \ref{table:abtest}, we perform ablation studies on the context-aware features by replacing them with frame-wise inputs, while keeping the remaining model configuration identical to Model 1. Compared to the results without context-aware features in Table \ref{table:abtest}, Model 1 in Table \ref{results} shows across the board performance improvements with  PCC scores up and RMSE down by an average of 0.015 and 0.045, respectively, across the development sets.

\subsubsection{Generalization error across datasets}
 
In our benchmarks of two previously published works \cite{NISQA, reddy2021dnsmos}, we notice significant performance inconsistencies across our development datasets. As shown in Table \ref{results}, the NISQA model performs noticeably better on the NISQA test datasets, whereas the DNSMOS struggles most on them (i.e. NISQA shows the highest PCC at 0.891 on the NISQA FOR testset, but an average of around 0.702 on the two challenge testsets; DNSMOS scores below 0.586 PCC on both NISQA testsets). This raises concerns about model generalization across varied domains.

We conduct further ablation studies by removing the PSTN training data in the training of Model 1, the result of which is shown in Table \ref{table:abtest}. We notice the performance remains similar in the other three development sets but drops off significantly in the PSTN development set, where PCC decreases from 0.754 to 0.654.

Although a thorough discussion of generalization error \cite{abu2012learning} in MOS prediction is beyond the scope of this work, we highlight the fact that model performance can be highly dataset sensitive in this task, even for corpora with outwardly similar content.

\subsection{Convolutional Context-aware Transformer}

Moving on to our proposed models, the individual results for the three selected models are presented in Table \ref{results}, we note that despite Model 3 outperforming Model 1 and Model 2 on all development datasets, the ensemble prediction further improves  overall robustness in both PCC and RMSE.
Comparing to the work of NISQA and DNSMOS, we are able to achieve significantly better performance on the two challenge datasets, with PCC around 0.901 and RMSE around 0.452 for the Tencent corpus and PCC around 0.779 and RMSE around 0.559 for the PSTN corpus. However, NISQA retains the best performance on its own test corpus.

Finally, we present the results for the challenge evaluation test set in Table \ref{table:test_result}. The baseline model is a simplified version of the state-of-art NISQA model\cite{NISQA}. This model uses a mel-spectral input to LSTM layers followed by several fully connected layers. An average pooling layer then used to average the predicted frame-level MOS and output the final predicted MOS score. The metrics also include the epsilon-insensitive RMSE after a 3rd order polynomial monotonic mapping (RMSE*3rd) \cite{itut1401,mittag2019non} which takes into account the confidence interval of the individual MOS scores.

The results indicate that our proposed model shows better performance across all three datasets under three different evaluation metrics. Overall, our CCAT model, reaches a 0.697 PCC while the baseline model only reach 0.53 PCC, indicating that the our model has a higher correlation to subjective speech assessment ratings. The RMSE and RMSE*3rd over all eval datasets for our proposed model is also lower than the baseline model. The proposed model reaches a 0.57 RMSE and 0.401 RMSE*3rd compare to 0.768 RMSE and 0.497 RMSE*3rd from the baseline model. Therefore, we can conclude that our proposed CCAT model shows promising results in the robust prediction of MOS scores under varied conditions.

\begin{table}[]
\resizebox{\columnwidth}{!}{%
\begin{tabular}{lcccccc}
\hline
\multirow{2}{*}{Dataset} & \multicolumn{3}{c}{Baseline 1}                                                   & \multicolumn{3}{c}{CCATMOS (Ensemble)}                                                                       \\ \cmidrule(r){2-4}  \cmidrule(r){5-7}  
                         & PCC   & RMSE  & \begin{tabular}[c]{@{}c@{}}RMSE*3rd\end{tabular} & PCC            & RMSE           & \begin{tabular}[c]{@{}c@{}}RMSE*3rd \end{tabular} \\ \hline
Tecent Eval              & 0.881 & 0.624 & 0.550                                                             & \textbf{0.945} & \textbf{0.397} & \textbf{0.38}                                                    \\
TUB Eval                 & 0.348 & 1.094 & 0.649                                                            & \textbf{0.613} & \textbf{0.852} & \textbf{0.552}                                                   \\
MS Eval                  & 0.361 & 0.585 & 0.293                                                            & \textbf{0.532} & \textbf{0.459} & \textbf{0.271}                                                   \\
Average                  & 0.530  & 0.768 & 0.497                                                            & \textbf{0.697} & \textbf{0.570}  & \textbf{0.401}                                                   \\ \hline
\end{tabular}}
\caption{Results on the challenge evaluation test set (obtained from the challenge host).  Note that the ``average" result is calculated without taking the quantity of each test set into account. Also note the ``MS Eval" result released from the challenge does not match any of the evaluation test set specification, here we assume this is the PSTN eval dataset.}
\label{table:test_result}
\end{table}

\section{Conclusions} \label{conclusion}
Subjective quality assessment by human raters is the best method available for assessing speech quality. However, it is time consuming and costly to conduct. In this paper, we proposed a novel end-to-end non-intrusive speech quality assessment model called convolutional context-aware transformer network  or CCAT. The proposed model architecture provides an accurate MOS prediction across both real and simulated conditions. The overall performance of our model is further improved by using hyper-parameter tuning and model ensembling techniques.

We also highlight the issue of poor model generalization across MOS corpora. We observe this behavior in our model as well as several other models from recent literature trained on cross-domain dataset splits. We hypothesize that this is likely a byproduct of the limited scale and variety of current MOS prediction datasets and may be exacerbated by the inherently noisy nature of subjective MOS ratings. As future work, we believe the robustness of the proposed model could be further improved with a larger and more diverse datasets as well as methods for cross-domain regularization. 

\bibliographystyle{IEEEtran}

\bibliography{mybib}


\end{document}